Entire latex paper,  no figures

\documentstyle[12pt]{article}
\def\lg{{\mathchoice{~\raise.58ex\hbox{$<$}\mkern-14.8mu\lower.52ex\hbox{$>$}~}
	            {~\raise.58ex\hbox{$<$}\mkern-14.8mu\lower.52ex\hbox{$>$}~}
		    {\raise.59ex\hbox{{$\scriptscriptstyle <$}}\mkern-12.8mu%
		     \lower.01ex\hbox{{$\scriptscriptstyle >$}}}   {} 	}} 
\def\gl{{\mathchoice{~\raise.58ex\hbox{$>$}\mkern-12.8mu\lower.52ex\hbox{$<$}~}
                    {~\raise.58ex\hbox{$>$}\mkern-12.8mu\lower.52ex\hbox{$<$}~}
		    {\raise.62ex\hbox{{$\scriptscriptstyle >$}}\mkern-12.0mu%
		     \lower.05ex\hbox{{$\scriptscriptstyle <$}}}  {} 	}}   

\begin{document}
\vspace{2cm}

\begin{center}

{\Large \bf Transport Theory of Massless Fields} \\

\vspace{1cm}

Stanis\l aw Mr\' owczy\' nski\footnote{E-mail: MROW@FUW.EDU.PL} \\[5mm]
\it
So\l tan Institute for Nuclear Studies,\\
ul. Ho\. za 69, PL - 00-681 Warsaw, Poland \\
and Institute of Physics, Pedagogical University,\\
ul. Le\' sna 16, PL - 25-509 Kielce, Poland\\
\vspace{2cm}
\rm
\begin{minipage}{13cm}
\baselineskip=12pt

{\small  \qquad Using the Schwinger-Keldysh technique we discuss how to 
derive the transport equations for the system of massless quantum fields.
We analyse the scalar field models with quartic and cubic interaction 
terms. In the $\phi^4$ model the massive quasiparticles appear due to 
the self-interaction of massless bare fields. Therefore, the derivation 
of the transport equations strongly resembles that one of the massive 
fields, but the subset of diagrams which provide the quasiparticle mass 
has to be resummed. The kinetic equation for the finite width quasiparticles 
is found, where, except the mean-field and collision terms, 
there are terms which are absent in the standard Boltzmann equation. The 
structure of these terms is discussed. In the massless $\phi^3$ model the 
massive quasiparticles do not emerge and presumably there is no transport 
theory corresponding to this model. It is not surprising since the $\phi^3$ 
model is anyhow ill defined.}

\end{minipage}
\end{center}
\vspace{0.5cm}

PACS numbers: 05.20.Dd, 11.10.Wx

\vspace{1cm}
\begin{center}
{\it 2-nd February 1997 \\}
\end{center}

\baselineskip=14pt
\newpage

\section{Introduction}

\qquad Transport theory is a very convenient tool to study many-body 
nonequilibrium systems, nonrelativistic and relativistic as well. 
The kinetic equations which play a central role in the transport approach
can be usually derived by means of simple heuristic arguments similar
to those which were used by Ludwig Boltzmann over hundred years ago when
he introduced his famous equation. However, such arguments fail when one
studies a system of very complicated dynamics as the quark-gluon plasma 
governed by QCD. In spite of some efforts, the general structure of collision 
terms in the transport equations of the quark-gluon plasma remains unknown. 
In such a situation one has to refer to a formal scheme which allows to derive 
the transport equation directly from the underlying quantum field theory. 
The formal scheme is also needed to specify the limits of the kinetic approach.
Indeed, the derivation shows the assumptions and approximations which lead to 
the transport theory, and hence the domain of its applicability can be 
established.

The so-called Schwinger-Keldysh \cite{Sch61} formulation of the quantum field 
theory provides a very promising basis to derive the transport equation. 
Kadanoff and Baym \cite{Kad62} developed the technique for nonrelativistic 
quantum systems, which has been further generalized to relativistic ones
\cite{Bez72,Li83,Dan84,Cal88,Mro90,Mro94,Hen95,Boy96}. We mention here only 
the papers which went beyond the mean field or Vlasov approximation and 
provide a more or less systematic analysis of the collision terms. 

The treatment of the massless fields, which are crucial for the
gauge theories as QED or QCD, is particularly difficult when the 
transport equations are derived. Except the well known infrared divergences 
which plague the perturbative expansion, there is a specific problem 
of nonequilibrium massless fields. The inhomogeneities in the system cause 
the off-mass-shell propagation of particles and then the perturbative 
analysis of the collision terms appears hardly tractable. More specifically, 
it appears very difficult, if possible at all, to express the field 
self-energy as the transition matrix element squared and consequently we loose 
the probabilistic character of the kinetic theory. The problem is absent 
for the massive fields when the system is assumed homogeneous at the 
inverse mass or Compton scale. This is a natural assumption within 
the transport theory which anyway deals with the quantities averaged over 
a certain scale which can be identified with the Compton one.

The problem of the massless {\it nonequilibrium} fields has not been fully 
recognized in the existing literature. One has usually assumed, explicitly 
or implicitly, the on-mass-shell propagation or equivalently the ideal 
homogeneity of the system. We find such an approach unsatisfactory and suggest 
a systematic solution of the problem. The basic idea is rather obvious.

The fields which are massless in vacuum gain an effective mass in 
a medium due to the interaction. Therefore, the minimal scale at which 
the transport theory works is not an inverse bar mass, which is infinite 
for massless fields, but the inverse effective one.  The staring point of 
the perturbative computation should be no longer free fields but 
the interacting ones. In physical terms, we postulate existence of 
the massive quasiparticles and look for their transport equation.

At technical level, we begin with the lagrangian of the massless fields 
and make a formal trick which is well known in the quantum field theory 
at finite temperature, see, e.g., 
\cite{Wei74,Dol74,Bra90,Tay90,Ban91,Par92,Efr93,Wan96}. Namely, the auxiliary
mass term is added to the free lagrangian and then is subtracted 
due to a redefinition of the interaction one. As a result the subset of
diagrams which contributes to the mass, which is determined in a self
consistent way, is effectively resummed in the perturbative expansion.
A somewhat similar technique was applied in the kinetic theory in \cite{Boy96}.

In this paper we show how the suggested method works for the self-interacting
scalar fields. We discuss in detail the $\phi^3$ and $\phi^4$ models which
appear to be qualitatively different. We successfully derive the transport 
equations for the $\phi^4$ model and show why the method does not work for 
the $\phi^3$ case. Our discussion closely follows the scheme of derivation 
which was earlier developed for the massive fields: self-interacting scalar 
fields \cite{Mro90} and the spinor fields interacting with the scalar and 
vector ones \cite{Mro94}. 

The main steps of the derivation are the following. We define the contour 
Green function with the time arguments on the contour in a complex time 
plane. This function is a key element of the Schwinger-Keldysh approach. 
After discussing its properties and relevance for nonequilibrium systems, 
we write down the exact equations of motion i.e. the Dyson-Schwinger 
equations. Assuming the macroscopic quasi-homogeneity of the system, 
we perform the gradient expansion and the Wigner transformation. Then, 
the pair of Dyson-Schwinger equations are converted into the transport
and mass-shell equations both satisfied by the Wigner function. 
The later equation allows one to identify the initially introduced 
fictitious mass with the effective one generated by the interaction. 
We further perform the perturbative analysis showing how the Vlasov
terms and the collisional ones emerge. Finally we define the
distribution functions of standard probabilistic interpretation
and find the transport equations satisfied by these functions.

Throughout the article we use natural units where $\hbar = c = 1$. The 
signature of the metric tensor is $(+,-,-,-)$. As far as possible we keep
the convention of Bjorken and Drell \cite{Bjo64}.

\section{Preliminaries}

\qquad We consider the system of massless scalar fields with the lagrangian 
density of the form
\begin{equation}\label{lagran}
{\cal L}(x) = {1 \over 2}\partial^{\mu}\phi (x)\partial_{\mu}\phi (x)
- {g \over n!}\phi^n(x) \;,
\end{equation}
where $n$ equals 3 or 4. The renormalization counterterms are omitted in 
the lagrangian. We introduce an auxiliary {\it position dependent} mass 
$m_*(x)$ which can be treated as an external field. Specifically, 
we redefine the lagrangian as
$$
{\cal L}_m(x) = {1 \over 2}\partial^{\mu}\phi (x) \partial_{\mu}\phi (x)
- {1 \over 2}m_*^2(x) \phi^2(x) + {\cal L}_I(x)
$$
with the interaction term
$$
{\cal L}_I(x) = + {1 \over 2} m_*^2(x) \phi^2(x) 
- {g \over n!}\phi^n(x) \;. 
$$

The fields which satisfy the equation of motion
\begin{equation}\label{motion}
\bigr[ \partial^2 + m_*^2(x) \bigl] \phi (x) = 0 \;, 
\end{equation}
represent {\it free quasiparticles} with mass $m_*$. We observe that it 
is not {\it a priori} clear whether massive quasiparticles emerge due to 
the field self-interaction. It is even less clear whether the limit of 
free quasiparticles exist. As will be shown it is indeed the case for
the $\phi^4$ model, but not for the $\phi^3$ one.

We write down the energy-momentum tensor defined as 
$$
T^{\mu \nu}(x) = \partial^{\mu}\phi (x)\partial^{\nu}\phi (x)
- g^{\mu \nu}{\cal L}(x) \;.
$$
Subtracting the total derivative 
$$
{1 \over 4} \partial^{\mu}\partial^{\nu} \phi^2(x) - 
g^{\mu \nu}{1 \over 4} \partial^{\sigma}\partial_{\sigma}
\phi^2(x) \;,
$$ 
we get the energy-momentum tensor which for the free fields is of 
the form convenient for our purposes i.e.
\begin{equation}\label{tensor}
T^{\mu \nu}_0 (x)= - {1 \over 4} \phi (x)\buildrel \leftrightarrow
\over \partial^{\mu} \buildrel \leftrightarrow
\over \partial^{\nu} \phi (x)  \;. 
\end{equation}
The fields are assumed here to satisfy the equation of motion
(\ref{motion}).

\section{Green functions}

\qquad The central role in our considerations plays the contour Green 
function defined as 
$$
i\Delta (x,y) \buildrel \rm def \over 
= \langle  \tilde T \phi (x) \phi (y) \rangle \; 
$$
where $\langle ...\rangle $ denotes the ensemble average at time $t_0$ 
(usually identified with $-\infty $); $\tilde T$ is the time ordering 
operation along the directed contour shown in Fig.~1. The parameter 
$t_{max}$ is shifted to $+ \infty$ in the calculations. The time arguments 
are complex with an infinitesimal positive or negative imaginary part, 
which locates them on the upper or on the lower branch of the contour. 
The ordering operation is defined as
$$
\tilde T \phi (x) \phi (y) \buildrel \rm def \over = 
\Theta (x_0,y_0)\phi (x) \phi (y) +
\Theta (y_0,x_0)\phi (y) \phi (x) \;,
$$
where $\Theta (x_0,y_0)$ equals 1 if $x_0$ succeeds $y_0$ on the 
contour, and  equals 0 when $x_0$  precedes $y_0$. 

If the field is expected to develop a finite expectation value,
as it happens when the symmetry is spontaneously broken, the
contribution $\langle \phi (x)\rangle \langle \phi (y) \rangle$  
is subtracted in the right-hand-side of the equation defining the Green 
function, see e.g. \cite{Mro90,Mro94}. Then, one concentrates on the 
field fluctuations around the expectation values. Since 
$\langle \phi (x)\rangle$ is expected to vanish in the models
defined by the lagrangians (\ref{lagran}) we neglect this contribution
in the Green function definition. 

We also use four other Green functions with real time arguments:
$$
i\Delta^> (x,y) \buildrel \rm def \over 
= \langle  \phi (x) \phi (y) \rangle \;, 
$$
$$
i\Delta^<  (x,y) \buildrel \rm def \over = 
\langle  \phi (y) \phi (x) \rangle \;, 
$$
$$
i\Delta^c (x,y) \buildrel \rm def \over = 
\langle  T^c \phi (x) \phi (y) \rangle \;, 
$$
$$
i\Delta^a (x,y) \buildrel \rm def \over 
= \langle  T^a \phi (x) \phi (y) \rangle \;, 
$$
where $T^c (T^a)$ prescribes (anti-)chronological time ordering:
$$
T^c \phi (x) \phi (y) \buildrel \rm def \over = 
\Theta (x_0-y_0) \phi (x) \phi (y) +
\Theta (y_0-x_0) \phi (y) \phi c(x) \;,
$$
$$
T^a \phi (x) \phi (y) \buildrel \rm def \over = 
\Theta (y_0-x_0) \phi (x) \phi (y) +
\Theta (x_0-y_0) \phi (y) \phi (x) \;.
$$
These functions are related to the contour Green functions 
in the following manner:
$$
\Delta^c(x,y) \equiv \Delta (x,y) \;\; {\rm for} \;\;
x_0 , \; y_0 \;\; {\rm from \;\; the \;\; upper \;\; branch,} 
$$
$$
\Delta^a(x,y) \equiv \Delta (x,y) \;\; {\rm for} \;\;
x_0 , \; y_0 \;\; {\rm from \;\; the \;\; lower \;\; branch,} 
$$
\begin{eqnarray*}
\Delta^>(x,y) \equiv \Delta (x,y) \;\; {\rm for} \;\;
&x_0&  \; {\rm from \;\; the \;\; upper \;\; branch\;\; and \;\;} \\
&y_0& \;\; {\rm from \;\; the \;\; lower \;\; one,} 
\end{eqnarray*}
\begin{eqnarray*}
\Delta^<(x,y) \equiv \Delta (x,y) \;\; {\rm for} \;\;
&x_0& \;\; {\rm from \;\; the \;\; lower \;\; branch  \;\; and \;\;} \\
&y_0& \;\; {\rm from \;\; the \;\; upper \;\; one. } 
\end{eqnarray*}

One easily finds the identities which directly follow from the
definitions
\begin{equation}\label{ident-c}
\Delta^c(x,y) = \Theta (x_0 - y_0) \Delta^>(x,y) +
\Theta (y_0 - x_0) \Delta^<(x,y) \;, 
\end{equation}
$$
\Delta^a(x,y) = \Theta (y_0 - x_0) \Delta^>(x,y) +
\Theta (x_0 - y_0) \Delta^<(x,y) \;. 
$$

One also observes that 
$$
\Bigr(i\Delta^{\lg}(x,y) \Bigl)^{\dagger} =  i\Delta^{\lg}(x,y) \;, 
$$
$$
\Bigr(i\Delta^a(x,y) \Bigl)^{\dagger} =  i\Delta^c(x,y) \;, 
$$
where $\dagger$ denotes hermitian conjugation, {\it i.e.} complex 
conjugation with an exchange of the Green function arguments. 
Because the fields are real, the functions $i\Delta^{\lg}(x,y)$ satisfy
the relation
\begin{equation}\label{neutral}
\Delta^>(x,y)  = \Delta^<(y,x)  \;. 
\end{equation}

It appears convenient to introduce the retarded $(+)$ and advanced
$(-)$ Green functions 
\begin{equation}\label{ret}
\Delta^{\pm} (x,y) \buildrel \rm def \over = \pm \Bigr(
\Delta^>(x,y) - \Delta^<(x,y) \Bigl) \Theta (\pm x_0 \mp y_0) \;.
\end{equation}
One immediately finds the identity
\begin{equation}\label{ret-adv}
\Delta^+(x,y)  - \Delta^-(x,y) = \Delta^>(x,y)  - \Delta^<(x,y) \;.
\end{equation}

Let us now briefly discuss the physical interpretation of the
Green functions. The function $\Delta^c(x,y)$ describes 
the propagation of disturbance in which a single particle is 
added to the many-particle system in space-time point $y$ and 
then is removed from it in a space-time  point $x$. 
An antiparticle disturbance is propagated backward in time. The meaning of 
$\Delta^a(x,y)$ is analogous but particles are propagated backward
in time and antiparticles forward. In the zero density limit 
$\Delta^c(x,y)$ coincides with the Feynman propagator.

The physical meaning of functions $\Delta^>(x,y)$ and 
$\Delta^<(x,y)$ is more transparent when one considers the Wigner 
transform defined as
\begin{equation}\label{Wigner}
\Delta^{\lg}(X,p) \buildrel \rm def \over = \int d^4u e^{ipu}
\Delta^{\lg}(X+{1 \over 2}u,X-{1 \over 2}u) \;.
\end{equation}
Then, the free-field energy-momentum tensor (\ref{tensor}) 
averaged over ensemble can be expressed as
\begin{equation}\label{en-mom-free}
\langle T^{\mu \nu}_0(X) \rangle = 
\int {d^4p \over (2\pi)^4} p^{\mu} p^{\nu} i\Delta^<(X,p) \;.
\end{equation}
One recognizes the standard form of the energy-momentum tensor 
in the kinetic theory with the function $i\Delta^<(X,p)$ giving 
the density of particles with four-momentum $p$ in a space-time 
point $X$. Therefore, $i\Delta^<(X,p)$ can be treated as a quantum 
analog of the classical distribution function. Indeed, the function 
$i\Delta^<(X,p)$ is hermitian. However it is not positively definite 
and the probabilistic interpretation is only approximately valid. 
One should also observe that, in contrast to the classical distribution
functions, $i\Delta^<(X,p)$ can be nonzero for the off-mass-shell 
four-momenta. 

\section{Green Function Equations of Motion}

\qquad The Dyson-Schwinger equations satisfied by the contour Green 
function are
\begin{equation}\label{DS1}
\big[\partial_x^2 + m_*^2(x)\big] \Delta (x,y) = 
-\delta^{(4)}(x,y) + \int_C d^4x' \Pi(x,x')
\Delta(x',y) \;, 
\end{equation}
\begin{equation}\label{DS2}
\big[\partial_y^2 + m_*^2(y)\big] \Delta (x,y) = 
-\delta^{(4)}(x,y) + \int_C d^4x' \Delta(x,x')
\Pi(x',y) \;,
\end{equation}
where $\Pi(x,y)$ is the self-energy; the  integration  over  
$x'_0$ is  performed  on  the contour and the function 
$\delta^{(4)}(x,y)$ is defined on the contour as 
\begin{displaymath}
\delta^{(4)}(x,y) = \left\{ \begin{array}{ccl} 
\delta^{(4)}(x-y) \;\;\; & {\rm for} &\;\; x_0 \;, \; y_0 \;\; 
{\rm from \;\; the \;\; upper \;\; branch,} \\
0 \;\;\;\;\;\;\; & {\rm for} &\;\; x_0 \;, \; y_0 \;\; 
{\rm from \;\; the \;\; different \;\; branches,} \\ 
-\delta^{(4)}(x-y) \;\;\; & {\rm for} & \;\; x_0 \;,\; y_0 \;\; 
{\rm from \;\; the \;\; lower \;\; branch.} \end{array} \right. 
\end{displaymath}

Let us split the self-energy into three parts as 
$$
\Pi (x,y) = \Pi_{\delta}(x)\delta^{(4)}(x,y) + \Pi^>(x,y) \Theta (x_0,y_0)
+ \Pi^<(x,y) \Theta (y_0,x_0) \;. 
$$
As we shall see later, $\Pi_{\delta}$ provides a dominant contribution
to the mean-field while $\Pi^{\gl}$ determines the collision terms in the
transport equations. Therefore, we call $\Pi_{\delta}$ the mean-field
self-energy and $\Pi^{\gl}$ the collisional self-energy.

With the help of the retarded and advanced Green functions 
(\ref{ret}) and the retarded and advanced self-energies defined in an 
analogous way, the equations (\ref{DS1}) and (\ref{DS2}) can be 
rewritten as  
\begin{eqnarray}\label{DSgl1}
\big[ \partial^2_x + m_*^2(x) &-& \Pi_{\delta}(x)\big ] 
\Delta^{\gl}(x,y) 
\nonumber \\ 
&=& \int d^4x' \Bigr[ \Pi^{\gl }(x,x') \Delta^{-}(x',y)+
\Pi^{+}(x,x') \Delta^{\gl}(x',y) \Bigl]  \;,
\end{eqnarray}
\begin{eqnarray}\label{DSgl2}
\big[ \partial^2_y  + m_*^2 &-& \Pi_{\delta}(y) \big] 
\Delta^{\gl }(x,y) 
\nonumber \\ 
&=&\int d^4x' \Bigr[ \Delta^{\gl}(x,x') \Pi^{-}(x',y)+
\Delta^{+}(x,x') \Pi^{\gl}(x',y)\Bigl]  \;, 
\end{eqnarray}
where all time integrations run from $- \infty$ to $+ \infty$.

Let us also write down the equations satisfied by the functions
$\Delta^{\pm}$ 
\begin{eqnarray}\label{DSpm1}
\big[\partial^2_x + m_*^2(x) &-& \Pi_{\delta}(x)\big] 
\Delta^{\pm}(x,y) \nonumber \\ 
&=& -\delta^{(4)}(x-y)+
\int d^4x'  \Pi^{\pm}(x,x') \Delta^{\pm}(x',y) \;,
\end{eqnarray}
\begin{eqnarray}\label{DSpm2}
\big[\partial^2_y + m_*^2(y) &-& \Pi_{\delta}(y)\big] 
\Delta^{\pm}(x,y)  \nonumber \\
&=& -\delta^{(4)}(x-y) + 
\int d^4x'  \Delta^{\pm}(x,x') \Pi^{\pm}(x',y) \;. 
\end{eqnarray}

\section{Towards Transport Equations}

\qquad The transport equations are derived under the assumption that 
the Green functions and the self-energies depend weakly on the sum 
of their arguments, and that they are significantly different from zero 
only when the difference of their arguments is close to zero. To express 
these properties it is convenient to define a new set of variables as
$$
\Delta (X,u)  \equiv \Delta (X+{1 \over 2}u,X-{1 \over 2}u) \;.
$$
For homogeneous systems, the dependence on $X = (x+y)/2$ drops out entirely
due to the translational invariance and $\Delta (x,y)$ depends only on 
$u = x-y$. For weakly inhomogeneous, or quasihomogeneous systems, 
the Green functions and self-energies are assumed to vary slowly with
$X$. We additionally assume that the Green functions and self-energies 
are strongly {\it peaked} near $u = 0$. The effective mass $m_*(x)$ is 
simply assumed to be weakly dependent on $x$.

We will now convert the equations (\ref{DSgl1}, \ref{DSgl2}) into transport 
equations by implementing the above approximation and performing the 
Wigner transformation (\ref{Wigner}) for all Green functions and 
self-energies. This is done using the following set of translation rules
which can be easily derived: 
\begin{eqnarray*}
\int d^4x' f(x,x') g(x', y) & \longrightarrow &
f(X,p)g(X,p) \\ 
& + & {i \over 2} \biggr[
{\partial f(X,p) \over \partial p_{\mu}}
{\partial g(X,p) \over \partial X^{\mu}} - 
{\partial f(X,p) \over \partial X^{\mu}}
{\partial g(X,p) \over \partial p_{\mu}} \biggl] \;, \\
h(x) g(x, y) & \longrightarrow &
h(X)g(X,p) -  {i \over 2} {\partial h(X) \over \partial X^{\mu}}
{\partial g(X,p) \over \partial p_{\mu}}\;, \\
h(y) g(x, y) & \longrightarrow &
h(X)g(X,p) +  {i \over 2} {\partial h(X) \over \partial X^{\mu}}
{\partial g(X,p) \over \partial p_{\mu}}\;,  \\
\partial^{\mu}_x f(x,y) & \longrightarrow &
(-ip^{\mu} + {1 \over 2} \partial^{\mu})f(X,p) \;, \\
\partial^{\mu}_y f(x,y) & \longrightarrow &
(ip^{\mu} + {1 \over 2} \partial^{\mu})f(X,p) \;. 
\end{eqnarray*}
Here $ X \equiv (x + y)/2$, $\;\partial^{\mu} \equiv {\partial \over
\partial X_{\mu}}$ and the functions $f(x,y)$ and $g(x,y)$ satisfy the 
assumptions discussed above. The function $h(x)$ is assumed to be weakly 
dependent on $x$.

Applying these translation rules to eqs.~(\ref{DSgl1}, \ref{DSgl2}), 
we obtain
\begin{eqnarray}\label{DSgl3}
\Bigr[ {1 \over 4} \partial^2  - 
ip^{\mu} \partial_{\mu} - p^2  &+&  m_*^2(X) - \Pi_{\delta}(X) 
 - {i \over 2} \partial_{\mu}
\Big( m_*^2(X) - \Pi_{\delta}(X) \Big) \partial^{\mu}_p \Bigl] 
\Delta^{\gl }(X,p)  \nonumber \\
& = & \Pi^{\gl }(X,p) \Delta^{-}(X,p)+
      \Pi^{+}(X,p) \Delta^{\gl }(X,p) \nonumber \\
& + & {i \over 2} \Big\{ \Pi^{\gl}(X,p),\,\Delta^{-}(X,p) \Big\}
  +   {i \over 2} \Big\{ \Pi^{+}(X,p),\, \Delta^{\gl}(X,p) \Big\} \;,
\end{eqnarray}
\begin{eqnarray}\label{DSgl4}
\Bigr[ {1 \over 4} \partial^2 +
ip^{\mu} \partial_{\mu} - p^2 &+&  m_*^2(X) - \Pi_{\delta}(X) 
 + {i \over 2} \partial_{\mu}
\Big( m_*^2(X) - \Pi_{\delta}(X) \Big) \partial^{\mu}_p \Bigl] 
\Delta^{\gl }(X,p)   \nonumber \\
& = & \Delta^{\gl }(X,p) \Pi^{-}(X,p)
    + \Delta^{+}(X,p) \Pi^{\gl}(X,p) \nonumber \\
& + & {i \over 2} \Big\{ \Delta^{\gl}(X,p),\, \Pi^{-}(X,p) \Big\}
  +   {i \over 2} \Big\{ \Delta^{+}(X,p),\, \Pi^{\gl}(X,p) \Big\} \;, 
\end{eqnarray}
where we have introduced the Poisson-like bracket defined as
$$
\Big\{ C(X,p),\, D(X,p) \Big\} \equiv
{\partial C(X,p) \over \partial p_{\mu}}
{\partial D(X,p) \over \partial X^{\mu}} - 
{\partial C(X,p) \over \partial X^{\mu}}
{\partial D(X,p) \over \partial p_{\mu}} \;.
$$

The kinetic theory deals only with averaged system characteristics.
Thus, one usually assumes that the system is homogeneous on a scale 
of the Compton wave length of the quasiparticles. In other words, 
the characteristic length of inhomogeneities is assumed to be much 
larger than the inverse mass of quasiparticles Therefore, we impose 
the condition
\begin{equation}\label{quasipar}
\Big\vert \Delta^{\gl }(X,p) \Big\vert \gg 
\Big\vert {1 \over m_*^2}\partial^2 
\Delta^{\gl}(X,p) \Big\vert \;,
\end{equation}
which leads to {\it the quasiparticle approximation}. As discussed in the
next section and in the Appendix, the requirement (\ref{quasipar}) renders 
the off-shell contributions to the Green functions $\Delta^{\gl}$ negligible. 
Thus, we deal with the quasiparticles having on-mass-shell momenta.
Unfortunately, the assumption (\ref{quasipar}) cannot be applied to massless 
particles and for this reason we have introduced the effective mass $m_*$. 

Let us now take the difference and the sum of eqs.~(\ref{DSgl3}) and 
(\ref{DSgl4}), where the $\partial^2$ terms have been neglected due 
to the quasiparticle approximation (\ref{quasipar}). Then, one gets 
\begin{eqnarray}\label{trans}
\Bigr[p^{\mu} \partial_{\mu} + {1 \over 2} \partial_{\mu}
\big( m_*^2(X) &-& \Pi_{\delta}(X) \big) \partial^{\mu}_p \Bigl] 
\Delta^{\gl }(X,p)  \nonumber \\
& = & {i \over 2} \Big( \Pi^>(X,p) \Delta^<(X,p) -
      \Pi^< (X,p) \Delta^> (X,p) \Big) \nonumber \\
& - & {1 \over 4} 
\Big\{ \Pi^{\gl}(X,p), \Delta^+(X,p) + \Delta^-(X,p) \Big\}  \nonumber \\
& - &  {1 \over 4} 
\Big\{ \Pi^+(X,p) + \Pi^+(X,p),\, \Delta^{\gl}(X,p) \Big\} \;,
\end{eqnarray}
\begin{eqnarray}\label{mass}
\Bigr[ -  p^2  &+&  m_*^2(X) - \Pi_{\delta}(X) \Bigl] 
\Delta^{\gl }(X,p)  \nonumber \\
& = & {1 \over 2} \Big( \Pi^{\gl }(X,p) 
\big( \Delta^{+}(X,p)+ \Delta^{-}(X,p) \big) 
      + \big( \Pi^{+}(X,p) + \Pi^{-}(X,p) \big) 
\Delta^{\gl }(X,p) \Big) \nonumber \\
& + & {i \over 4} \Big\{ \Pi^>(X,p),\,\Delta^<(X,p) \Big\} 
 -  {i \over 4} \Big\{ \Pi^<(X,p),\, \Delta^>(X,p) \Big\} \;,
\end{eqnarray}
where we have used the identity (\ref{ret-adv}) applied to the Green
functions and self-energies. 

One recognizes eq.~(\ref{trans}) as a transport equation while 
eq.~(\ref{mass}) as a so-called mass-shell equation. We will write 
down these equation in a more compact way. From the definition (\ref{ret}) 
one finds that
\begin{eqnarray}\label{ret1}
\Delta^{\pm}(X,p) = \pm {1 \over 2} \Bigr(  
\Delta^>(X,p) & - & \Delta^<(X,p) \Bigl)  \nonumber \\
& + & {1 \over 2 \pi i} {\rm P}
\int d\omega ' {\Delta^>(X,\omega ', {\bf p})
- \Delta^< (X,\omega ', {\bf p}) \over \omega -\omega '} \;.
\end{eqnarray}
The first term in the r.h.s is antihermitian while the second one
is hermitian. Thus, we introduce
\begin{eqnarray}\label{Imdelta}
{\rm Im} \Delta^{\pm}(X,p) \equiv \pm {1 \over 2i} 
\Bigr( \Delta^>(X,p) -  \Delta^<(X,p) \Bigl) \;,
\end{eqnarray}
\begin{eqnarray}\label{Redelta}
{\rm Re} \Delta^{\pm}(X,p) \equiv {1 \over 2 \pi i} {\rm P}
\int d\omega ' {\Delta^>(X,\omega ', {\bf p})
- \Delta^< (X,\omega ', {\bf p}) \over \omega -\omega '} \;.
\end{eqnarray}

With the help of (\ref{Redelta}) and analogous formulas
for $\Pi^{\pm}$ the equations (\ref{trans}, \ref{mass}) 
can be rewritten as
\begin{eqnarray}\label{trans1}
\Big\{ p^2 - m_*^2(X) + \Pi_{\delta}(X) &+& {\rm Re} \Pi^+(X,p),\, 
\Delta^{\gl }(X,p) \Big\}   \nonumber \\
& = & i \Big( \Pi^>(X,p) \Delta^<(X,p) -
      \Pi^< (X,p) \Delta^> (X,p) \Big) \nonumber \\
& - & \Big\{ \Pi^{\gl}(X,p),\, {\rm Re} \Delta^+(X,p) \Big\}  \;,
\end{eqnarray}
\begin{eqnarray}\label{mass1}
\Bigr[ p^2 - m_*^2(X) &+& \Pi_{\delta}(X) + {\rm Re} \Pi^{+}(X,p) 
\Bigl] \Delta^{\gl }(X,p)  
= - \Pi^{\gl }(X,p) {\rm Re} \Delta^{+}(X,p) 
\nonumber \\
&-& {i \over 4} \Big\{ \Pi^>(X,p),\,\Delta^<(X,p) \Big\} 
 +  {i \over 4} \Big\{ \Pi^<(X,p),\, \Delta^>(X,p) \Big\} \;.
\end{eqnarray}

In the case of fields with the finite bare mass the gradient terms
in the right-hand-sides of eqs.~(\ref{trans1}, \ref{mass1}) are
{\it small} \cite{Mro90,Mro94} and are usually neglected. When 
the bare fields are massless, as those studied here, there is no
reason to neglect the gradient terms. The equation analogous to 
(\ref{trans1}) was earlier derived in \cite{Kad62,Bez72}. 

It appears useful to write down the transport and mass-shell
equations satisfied by the retarded and advanced Green functions.
Starting with eqs.~(\ref{DSpm1},\ref{DSpm2}) one finds
\begin{eqnarray}\label{trans-pm}
\Big\{ p^2 - m_*^2(X) + \Pi_{\delta}(X) + \Pi^{\pm}(X,p), 
\Delta^{\pm}(X,p) \Big\}  
 = 0  \;,
\end{eqnarray}
\begin{eqnarray}\label{mass-pm}
\Bigr[ p^2  -  m_*^2(X) + \Pi_{\delta}(X) + \Pi^{\pm}(X,p) \Bigl] 
\Delta^{\pm}(X,p)  =   1 \;.
\end{eqnarray}
We observe that the gradient terms drop out entirely in eq.~(\ref{mass-pm}).
Nevertheless the equation holds within the first order of gradient expansion. 
Due to the absence of the gradients, eq.~(\ref{mass-pm}) can be immediately 
solved as
\begin{eqnarray}\label{Green-pm}
\Delta^{\pm}(X,p) = 
{1 \over  p^2  -  m_*^2(X) + \Pi_{\delta}(X) + \Pi^{\pm}(X,p)}  \;.
\end{eqnarray}
One notices that $\Delta^{\pm}$ of the form (\ref{Green-pm}) solves
not only eq.~(\ref{mass-pm}) but eq.~(\ref{trans-pm}) as well. Indeed,
any function $f$ of $K$ satisfies the equation $\{ K, f(K) \} = 0$.

The real and imaginary parts of $\Delta^{\pm}(X,p)$, which are needed
in our further considerations, are
\begin{eqnarray}\label{Re-pm}
{\rm Re} \Delta^{\pm}(X,p) = 
{ p^2  -  m_*^2(X) + \Pi_{\delta}(X) + {\rm Re}\Pi^+(X,p)
\over \big( p^2  -  m_*^2(X) + \Pi_{\delta}(X) + {\rm Re}\Pi^+(X,p) \big)^2
+ \big({\rm Im}\Pi^+(X,p) \big)^2} \;,
\end{eqnarray}
\begin{eqnarray}\label{Im-pm}
{\rm Im} \Delta^{\pm}(X,p) = 
{ \pm {\rm Im}\Pi^+(X,p)
\over \big( p^2  -  m_*^2(X) + \Pi_{\delta}(X) + {\rm Re}\Pi^+(X,p) \big)^2
+ \big({\rm Im}\Pi^+(X,p) \big)^2} \;.
\end{eqnarray}

\section{Free quasiparticles}

\qquad Before further analysis the equations obtained in the previous 
section we consider here a very important limit which corresponds to the 
free quasiparticles. Specifically, we assume that 
$\Pi_{\delta} = \Pi^{\gl }= 0$. Then, the equations 
(\ref{trans1}, \ref{mass1}) read
\begin{eqnarray}\label{trans0}
\Bigr[p^{\mu} \partial_{\mu} + 
{1 \over 2} \partial_{\mu} m_*^2(X) \partial^{\mu}_p \Bigl] 
\Delta^{\gl}_0(X,p) = 0 \;,
\end{eqnarray}
\begin{equation}\label{mass0}
\big[ p^2 - m_*^2 \big] \Delta^{\gl}_0(X,p) = 0 \;.
\end{equation}
Although the quasiparticles are assumed to be free, the transport equation
is of the Vlasov not of the free form. This is a simple consequence of 
the $X-$dependence of the effective mass. 

Due to eq.~(\ref{mass0}) $\Delta^{\gl}_0(X,p)$ is proportional to
$\delta (p^2 - m_*^2)$, and consequently {\it free quasiparticles are always 
on mass-shell.} If the quasiparticle approximation (\ref{quasipar}) is 
{\it not} applied, the mass-shell equation gets the form
\begin{eqnarray*}
\Bigr[ {1 \over 4} \partial^2  - p^2  
+  m_*^2(X)  \Bigl] \Delta^{\gl}_0(X,p) = 0 \;,
\end{eqnarray*}
and the off-shell contribution to the Green function $\Delta^{\gl}_0$ 
is nonzero. A detailed discussion of the quasiparticle approximation is
given in the Appendix.

We also discuss the (anti-)chronological Green functions 
$\Delta^{c(a)}_0$ in the limit of free quasiparticles. One easily finds 
their equations of motion as
$$
\Bigr[p^{\mu} \partial_{\mu} + 
{1 \over 2} \partial_{\mu} m_*^2(X) \partial^{\mu}_p \Bigl] 
\Delta^{c}_0(X,p) = 0 \;,  
$$
$$
\big[p^2 - m_*^2 \big] \Delta^{c}_0(X,p) = 1 \;. 
$$
For the antichronological function $\Delta^{a}$, the right hand side 
of the mass-shell equation equals $-1$ instead of $+1$. The solution of 
these equation can be written as
$$
\Delta^c_0(X, p) = {1 \over p^2 -m_*^2 + i0^+} + 
\Theta (-p_0) \Delta^>_0(X,p) +\Theta (p_0) \Delta^<_0(X,p) \;, 
$$
where $\Delta^{\gl}_0(X,p)$ is assumed to satisfy 
eqs.~(\ref{trans0}, \ref{mass0}). It is worth mentioning that any function 
which depends on $(X,p)$ through $(p^2 - m_*^2)$ solves the Vlasov equation 
(\ref{trans0}). $\Delta^c_0(X, p)$ obeys the initial condition of the 
standard Feynman propagator. It also satisfies the relation (\ref{ident-c}).

The antichronological Green function is 
$$
\Delta^a_0(X, p) = {-1 \over p^2 -m_*^2 - i0^+} + 
\Theta (p_0) \Delta^>_0(X,p) + \Theta (-p_0) \Delta^<_0(X,p) \;.
$$

Knowing $\Delta^c_0$ and $\Delta^a_0$ one immediately gets the retarded
and advanced functions 
\begin{equation}\label{pm0}
\Delta^{\pm}_0(X, p) = {1 \over p^2 -m_*^2 \pm ip_00^+} \;,
\end{equation}
which obey the respective initial conditions. Confronting the expressions
of $\Delta^{\pm}$ for free (\ref{pm0}) and interacting quasiparticles
(\ref{Green-pm}), one finds that 
\begin{equation}\label{free-limit}
{\rm Im}\Pi^+(X,p) = - {\rm Im} \Pi^-(X,p) =
\left\{ \begin{array}{ccl} 
0^+ \;\;\; & {\rm for} &\;\; p_0 > 0 \;, \\
0^- \;\;\; & {\rm for} & \;\; p_0 < 0 \;, \\
\end{array} \right. 
\end{equation}
in the limit of free quasiparticles.

It appears useful to express the Green functions $\Delta^{\gl}_0$ 
through the distribution function $f_0$ as 
\begin{eqnarray}\label{def1-f0}
\Theta (p_0) i \Delta^<_0(X,p) = \Theta (p_0)
\;2\pi\delta (p^2-m_*^2)f_0(X,{\bf p}) =
{\pi \over E_p}\delta (E_p - p_0) f_0(X,{\bf p}) \;,
\end{eqnarray}
where $E_p \equiv \sqrt{{\bf p}^2 + m_*^2}\,$. This equation
should be treated as a definition of $f_0$.

Due to the relation (\ref{neutral}) we have
$$
\Delta^<(X,p) = \Delta^>(X,-p) \;,
$$
and consequently 
\begin{eqnarray}\label{def2-f0}
\Theta (p_0) i \Delta^>_0(X,-p) 
 =  \Theta (p_0) \;2\pi\delta (p^2-m_*^2 ) f_0(X,{\bf p}) 
 = {\pi \over E_p}\delta (E_p - p_0) f_0(X,{\bf p}) \;.
\end{eqnarray}

In that way we express the positive energy part of $\Delta^<_0(X,p)$
and the negative energy part of $\Delta^>_0(X,p)$ through $f_0(X,{\bf p})$.
We extend these expressions to the whole energy domain using 
the identity (\ref{ret-adv}). With the help of the explicit form
of the retarded and advanced functions (\ref{pm0}) we get the formula
\begin{equation}\label{gr-sm}
i\Delta^>_0(X,p)-i\Delta^<_0(X,p)= 2\pi \delta (p^2 - m_*^2 )
\bigr( \Theta (p_0) - \Theta (-p_0) \bigl) \;,
\end{equation}
which is discussed in detail in the next section.

Combining eqs. (\ref{def1-f0}, \ref{def2-f0}) with (\ref{gr-sm})
one finds the desired expression of the Green functions $\Delta^{\gl}_0$ 
in terms of the distribution function $f_0$. Namely,
\begin{eqnarray}\label{sm-f0}
i \Delta^<_0(X,p)  =  {\pi \over E_p}\delta (E_p - p_0) f_0(X,{\bf p}) 
+ {\pi \over E_p}\delta (E_p + p_0) \big( f_0(X,-{\bf p}) + 1 \big) \;,
\end{eqnarray}
\begin{eqnarray}\label{gr-f0}
i \Delta^>_0(X,p)  =  
{\pi \over E_p}\delta (E_p - p_0) \big( f_0(X,{\bf p}) + 1 \big)
+ {\pi \over E_p}\delta (E_p + p_0) f_0(X,-{\bf p}) \;.
\end{eqnarray}

When the system is in thermodynamical equilibrium the distribution
functions reads 
\begin{equation}\label{equi}
f_0^{\rm eq}({\bf p}) = { 1 \over e^{\beta^{\mu}p_{\mu}} - 1} \;,
\end{equation}
where $\beta^{\mu} \equiv u^{\mu}/T$ with $u^{\mu}$ being the hydrodynamical
velocity and $T$ the temperature. In the {\it local} equilibrium the
two parameters are $X-$dependent. 

\section{Spectral function}

\qquad In this section we introduce one more function which appears 
useful in the analysis of the interacting systems. The spectral function $A$
is defined as
$$
A(x,y) \buildrel \rm def \over  = i\Delta^>(x,y)  -  i\Delta^<(x,y) \;.
$$
Thus, 
$$
A(x,y) = \langle [ \phi (x),  \phi (y)] \rangle \;, 
$$
where $[\phi (x),  \phi (y)]$  denotes the field commutator.

Due to the equal time commutation relations
$$
[ \phi (t,{\bf x}),  \phi (t,{\bf y})] = 0 \;,\;\;\;\;\;\;\;
[\dot \phi (t,{\bf x}), \phi (t,{\bf y})] = -i \delta ^{(3)}
({\bf x}-{\bf y}) \;, 
$$
with the dot denoting the time derivative, the Wigner transformed
spectral function satisfies the two identities 
\begin{equation}\label{spec-id}
\int {dp_0 \over 2\pi } \,  A(X,p) = 0 \;, \;\;\;\;\;\;\;
\int {dp_0 \over 2\pi } \, p_0 \, A(X,p) =  1 \;.
\end{equation}
One also sees that (cf. eq.~(\ref{Imdelta}))
\begin{equation}\label{spec-Im}
A(X,p) = \mp 2 \, {\rm Im} \Delta^{\pm}(X,p) \;.
\end{equation}
Finally, we observe that the identity (\ref{neutral}), which holds for
the real fields, provides the relation
$$
A(X,p) = - A(X,-p) \;.
$$
From the transport and mass-shell equations (\ref{trans1}, \ref{mass1}) 
one immediately finds the equations for $A(X,p)$ as
\begin{equation}\label{trans-spec}
\Big\{ p^2 - m_*^2(X) + \Pi_{\delta}(X) + {\rm Re} \Pi^+(X,p), \,
A(X,p) \Big\} = 2\Big\{ {\rm Im}\Pi^+(X,p),\,{\rm Re}\Delta^+(X,p) \Big\} \;,
\end{equation}
\begin{equation}\label{mass-spec}
\Bigr[ p^2 - m_*^2(X) + \Pi_{\delta}(X) + {\rm Re} \Pi^+(X,p) 
\Bigl] A(X,p)  
= 2 \, {\rm Im}\Pi^+(X,p) \, {\rm Re} \Delta^+(X,p) \;.
\end{equation}

Substituting ${\rm Re} \Delta^+(X,p)$ from eq.~(\ref{Re-pm}) into 
the algebraic equation (\ref{mass-spec}) we find its solution as
\begin{eqnarray}\label{spec}
A (X,p) = { 2 {\rm Im}\Pi^+(X,p)
\over \big( p^2  -  m_*^2(X) + \Pi_{\delta}(X) 
+ {\rm Re}\Pi^+(X,p) \big)^2 + \big({\rm Im}\Pi^+(X,p) \big)^2} \;.
\end{eqnarray}
Then, one easily shows that the function of the form (\ref{spec}) solves
eq.~(\ref{trans-spec}) as well. In fact, the spectral function (\ref{spec})
can be found directly from eq.~(\ref{Re-pm}) due to the relation
(\ref{spec-Im}).

The spectral function of the free quasiparticles can be, obviously,
obtained from (\ref{spec}) but the limit should be taken with care.
We firstly write the spectral function (\ref{spec}) as 
\begin{eqnarray*}
A (X,p) &=& 
{i \over  p^2  -  m_*^2(X) + \Pi_{\delta}(X) + {\rm Re} \Pi^+(X,p) 
+ i {\rm Im}\Pi^+(X,p) } \\
&-& {i \over  p^2  -  m_*^2(X) + \Pi_{\delta}(X) + {\rm Re} \Pi^+(X,p) 
- i {\rm Im}\Pi^+(X,p) } \;.
\end{eqnarray*}
Then we take the limit $\Pi \rightarrow 0$ keeping in mind the 
condition (\ref{free-limit}). Using the well-known identity
$$
{1 \over x \pm i 0^+} = {\rm P}{1 \over x} \mp i \pi \delta (x) \;,
$$
we get the spectral function of noninteracting quasiparticles as
\begin{eqnarray}\label{spec-free}
A_0 (X,p) = 2\pi \delta (p^2 - m_*^2 )
\bigr( \Theta (p_0) - \Theta (-p_0) \bigl) \;,
\end{eqnarray}
which, of course, coincides with (\ref{gr-sm}).

Let us also consider a specific approximate form of the spectral function.
If the condition
$$
{\bf p}^2  +  m_*^2(X) - \Pi_{\delta}(X) 
- {\rm Re}\Pi^+(X,p) \gg \vert {\rm Im}\Pi^+(X,p) \vert
$$
is satisfied, $A$ as a function of $p_0$ is close to zero everywhere except 
two narrow regions around $p_0= \pm E_{\pm}$ ($E_{\pm} > 0$) which solve 
the equations
$$
E_{\pm}^2 (X,{\bf p}) = {\bf p}^2  +  m_*^2(X) - \Pi_{\delta}(X) 
- {\rm Re}\Pi^+(X,p_0 = \pm E_{\pm}, {\bf p}) \;.
$$
Then, the function (\ref{spec}) can be approximated as
\begin{eqnarray}\label{spec-app}
A (X,p) &=& {1 \over E_+(X,{\bf p})} \:
{\Gamma_+(X,{\bf p}) \over \big(E_+(X,{\bf p})-p_0 \big)^2 + 
\Gamma_+^2(X,{\bf p})} \nonumber\\
&-& {1 \over E_-(X,{\bf p})} \:
{\Gamma_-(X,{\bf p}) \over \big(E_-(X,{\bf p})+p_0 \big)^2 + 
\Gamma_-^2(X,{\bf p})} \;,
\end{eqnarray}
with
\begin{equation}\label{width}
\Gamma_{\pm}(X,{\bf p}) \equiv \pm \:
{{\rm Im}\Pi^+(X,p_0 = \pm E_{\pm}, {\bf p}) \over 2E_{\pm} (X,{\bf p})} \;.
\end{equation}
One easily checks that the spectral function of the form (\ref{spec-app})
satisfies the sum rules (\ref{spec-id}). 

\section{Perturbative expansion}

\qquad As discussed in {\it e.g.} \cite{Dan84,Cal88,Hen90} the contour 
Green functions admit a perturbative expansion very similar to that 
known from  vacuum field theory with essentially the same Feynman rules.
However, the time integrations do not run from $-\infty $ to $+\infty $, 
but along the contour shown in Fig.~1. The right turning point of the
contour ($t_{max}$) must be above the largest time argument of the evaluated
Green function. In practice, $t_0$ is shifted to $-\infty $ and $t_{max}$ 
to $+\infty $. The second difference is the appearance of tadpoles, 
{\it i.e.} loops formed by single lines, which give zero contribution 
in the vacuum case. A tadpole corresponds to a Green function with two 
equal space-time arguments. Since the Green function  $\Delta(x,y)$ is 
not well defined for $x = y$ we ascribe the function $-i\Delta^<(x,x)$ 
to each tadpole. The rest of Feynman rules can be taken from the textbook 
of Bjorken and Drell \cite{Bjo64}.

In this section we consider the lowest-order contributions to the 
self-energies. It should be stressed that the Green functions, which 
are represented by the lines of the Feynman diagrams, correspond to those 
of free quasiparticles not of noninteracting fields. 

\subsection*{$\phi^4$ model}

\qquad The lowest-order contribution to the self-energy which is associated
with the graphs from Fig.~2 is
$$
\Pi (x,y) = \delta^{(4)}(x,y) 
\Big( m_*^2(x) - {ig\over 2} \Delta^<_0(x,x) \Big) \;,
$$
giving
\begin{equation}\label{pi-delta}
\Pi_{\delta}(x) = m_*^2(x) -{ig\over 2} \Delta^<_0(x,x) \;,
\end{equation}
and
$$
\Pi^>(x,y) = \Pi^<(x,y) = 0 \;.
$$

Substituting $\Delta^<_0$ given by eq.~(\ref{sm-f0}), where $\Delta^<_0$ 
is expressed through the distribution function, into (\ref{pi-delta}) one 
finds $\Pi_{\delta}$ as
$$
\Pi_{\delta}(x) = m_*^2(x) -{g \over 2} \int {d^3 p \over (2\pi )^3 2E_p}
\big[ 2 f_0(x,{\bf p}) + 1] \;.
$$
As seen the integral is quadratically divergent in the vacuum limit
when $f_0 \rightarrow 0$. This type of divergence, which appears due to
the zero-mode fluctuations, is well known in the field theory. 
We remove it away by subtracting the vacuum value from $\Pi_{\delta}$.
Thus one gets
\begin{equation}\label{pi-delta-4-f}
\Pi_{\delta}(x) = m_*^2(x) -{g \over 2} \int {d^3 p \over (2\pi )^3 E_p}
f_0(x,{\bf p}) \;.
\end{equation}

We compute $\Pi_{\delta}$ for the equilibrium system when the distribution 
function is given by eq.~(\ref{equi}) with $u^{\mu} = (1,0,0,0)$. 
For $T \gg m_*$ we get after elementary integration the well-known 
result, see e.g. \cite{Par92}, 
\begin{equation}\label{pi-delta-4-eq}
\Pi_{\delta}(x) = m_*^2(x) -{g T^2(x) \over 24} \;,
\end{equation}
where the temperature is $x-$dependent in the case of {\it local} 
equilibrium.

\subsection*{$\phi^3$ model}

\qquad The lowest-order contribution to the self-energy corresponding 
to the graphs from Fig.~3a and 3b is
$$
\Pi_a(x,y) = \delta^{(4)}(x,y)
\Big[ m_*^2 - {i \over 2}g^2  \int_C d^4x' 
\Delta_0(x',x) \Delta^<_0(x',x') \Big] \;. 
$$
Locating the argument $x$ on the upper branch of the contour, 
one finds
$$
\Pi_{\delta}(x)  = m_*^2 - {i \over 2} g^2 
\int  d^4x' \Bigr[ \Delta^c_0(x',x) - \Delta^<_0 (x',x) \Bigl] 
\Delta^<_0(x',x') \;, 
$$
where the time integration runs from $-\infty $ to $+\infty $. 
Observing that $\Delta^c - \Delta^< = \Delta^+$ we get
$$
\Pi_{\delta}(x) = m_*^2 - {i \over 2} g^2 \int  d^4x' 
\Delta^+_0(x',x) \Delta^<_0(x',x') \;. 
$$

Using the explicit form of $\Delta^+_0$ (\ref{pm0}) and expressing 
$\Delta^<_0$ through the distribution function, one finds
\begin{equation}\label{pi-delta-3-f}
\Pi_{\delta}(x) = m_*^2(x) - { g^2 \over 2} \int 
{d^4 x' d^4 p \over (2\pi )^4}\; {d^3 k \over (2\pi )^3 E_k} \;
 {e^{-ip(x'-x)} \over p^2 - m_*^2 + ip_00^+} \; f_0(x',{\bf k}) \;,
\end{equation}
where as previously we have subtracted the vacuum contribution.

In the case of {\it global} equilibrium, when $f_0(x,{\bf k})$ is 
independent of $x$, the integral from (\ref{pi-delta-3-f}) can be 
computed as
\begin{equation}
\Pi_{\delta} = m_*^2 + { g^2 \over 2 m_*^2} 
\int  {d^3 k \over (2\pi )^3 E_k} \; f_0({\bf k}) \;.
\end{equation}
In the limit $T \gg m_*$ one finally finds
\begin{equation}\label{pi-delta-3-eq}
\Pi_{\delta} = m_*^2 + {g^2 \over 24} \, {T^2 \over m_*^2} \;.
\end{equation}

The graph from Fig.~3c corresponds to 
$$
\Pi_b(x,y) = -{i\over 2}g^2 \Delta_0(x,y) \Delta_0(y,x) \;,
$$
and it gives
\begin{equation}\label{pi-><-3}
\Pi^{\gl}(x,y) = -{i\over 2}g^2 
\Delta^{\gl}_0(x,y) \Delta^{\lg}_0(y,x) \;. 
\end{equation}

Substituting $\Delta^{\gl}$ expressed through the distribution 
function to eq.~(\ref{pi-><-3}) one finds
\begin{eqnarray}\label{pi->-3-f}
\Pi^>(X,p) &=& {i\over 2}g^2 \int {d^3 k \over (2\pi )^3 2E_k} \,
{d^3 q \over (2\pi )^3 2E_q} \, (2\pi)^4 
\delta^{(3)} ({\bf p} + {\bf q} - {\bf k}) \nonumber\\
&&\Big[ \;
\delta (p_0 - E_k - E_q) \, f_0(X,{\bf k}) \, f_0(X,-{\bf q}) \nonumber\\
&&+\;\delta (p_0 - E_k + E_q) \, f_0(X,{\bf k}) \, \big(f_0(X,{\bf q}) + 1\big) 
                                                       \nonumber\\
&&+\;\delta (p_0 + E_k - E_q) \, \big( f_0(X,-{\bf k}+1) \, f_0(X,-{\bf q})
                                                        \nonumber\\
&&+\;\delta (p_0 + E_k + E_q) \, \big( f_0(X,-{\bf k})+1\big) \,
                            \big( f_0(X,{\bf q}) + 1 \big) \; \Big] \;,
\end{eqnarray}
\begin{eqnarray}\label{pi-<-3-f}
\Pi^<(X,p) &=& {i\over 2}g^2 \int {d^3 k \over (2\pi )^3 2E_k} \,
{d^3 q \over (2\pi )^3 2E_q} \, (2\pi)^4 
\delta^{(3)} ({\bf p} + {\bf q} - {\bf k}) \nonumber\\
&&\Big[ \;
\delta (p_0 - E_k - E_q) \, \big( f_0(X,{\bf k})+1\big) \,
\big( f_0(X,-{\bf q}) + 1 \big) \nonumber\\
&&+\;\delta (p_0 - E_k + E_q) \, \big( f_0(X,{\bf k})+ 1\big) \, f_0(X,{\bf q})  
                                                       \nonumber\\
&&+\;\delta (p_0 + E_k - E_q) \, f_0(X,-{\bf k}) \, \big( f_0(X,-{\bf q})+1\big)
                                                         \nonumber\\
&&+\;\delta (p_0 + E_k + E_q) f_0(X,-{\bf k}) f_0(X,{\bf q}) \Big] \;.
\end{eqnarray}
Let us observe here that due to the trivial kinematical reasons 
$\Pi^<(X,p) = \Pi^<(X,p) = 0$ for on-mass-shell four-momenta $p$ i.e.
when $p^2 = m_*^2$. 

We further compute $\Pi^<(X,p)$ for ${\bf p} = 0$. In the case of equilibrium 
when the distribution function is of the form (\ref{equi}) and, as
previously, $u^{\mu} = (1,0,0,0)$, one finds $\Pi^{\gl}(X,p)$ for 
$p = (\pm \, \omega, {\bf 0})$ with $\omega > 0$ as
\begin{eqnarray}\label{pi->-3-eq}
\Pi^>(X,\omega, {\bf 0}) &=& \Pi^<(X,- \omega, {\bf 0}) \nonumber \\
&=& - {ig^2\over 16 \pi} \,
\Theta(\omega - 2m_* ) \,
\sqrt{1 - {4m_*^2 \over \omega^2}} \,
{1 \over 
\Big( \exp \big( {\beta \omega \over 2}\big) - 1 \Big)^2 } \;,
\end{eqnarray}
\begin{eqnarray}\label{pi-<-3-eq}
\Pi^<(X,\omega, {\bf 0}) &=& \Pi^>(X,-\omega, {\bf 0}) \nonumber \\
&=& - {ig^2\over 16 \pi} \,
\Theta( \omega - 2m_* ) \,
\sqrt{1 - {4m_*^2 \over \omega^2}} \,
{ \exp \big( \beta  \omega  \big)  \over 
\Big( \exp \big( {\beta \omega \over 2} \big) - 1 \Big)^2 } \;.
\end{eqnarray}
We also compute
\begin{eqnarray*}
{\rm Im} \Pi^+(X,\pm \, \omega, {\bf 0}) &=&
{1 \over 2i} \Big(
\Pi^>(X,\pm \, \omega, {\bf 0}) - \Pi^<(X,\pm \, \omega, {\bf 0}) \Big) \\
&=& \pm {g^2\over 16 \pi} \,
\Theta(\omega - 2m_* ) \,
\sqrt{1 - {4m_*^2 \over \omega^2}} \,
{1 \over  \exp \big( {\beta \omega \over 2}\big) - 1 } \;.
\end{eqnarray*}
One sees that the condition (\ref{free-limit}) is indeed satisfied by the 
perturbative self-energy when $g \rightarrow 0$ and $m_*$ is kept fixed.

\section{Quasiparticle mass}

\qquad The structure of eqs.~(\ref{trans1}, \ref{mass1}) motivates 
the definition of the quasiparticle mass as a solution of the equation
\begin{equation}\label{mstar}
\Pi_{\delta}(X) + {\rm Re} \Pi^{+}(X,p_0 = m_* , {\bf p} = 0) = 0 \;.
\end{equation}
The definition is not Lorentz invariant but statistical systems usually
break such an invariance. In the case of global equilibrium there is,
for example, a preferential reference frame related to the thermostat.

Let us now look for the explicit expression of $m_*$ in the lowest
nontrivial order in $g$ within the two models under consideration.

\subsection*{$\phi^4$ model}

\qquad The only nonvanishing contribution to the self-energy in the lowest 
order of perturbative expansion comes from $\Pi_{\delta}$, which is given by 
eq.~(\ref{pi-delta-4-f}). Therefore, the effective mass is a solution of 
the equation
\begin{equation}\label{eff-mass}
m_*^2(x) = {g \over 2} \int {d^3 p \over (2\pi )^3 E_p}
f_0(x,{\bf p}) \;.
\end{equation}
One should keep in mind that $E_p$ and $f_0$ depend on $m_*$.

For the equilibrium system with $T \gg m_*$ we immediately get
the well known result:
\begin{equation}\label{mass-4-eq}
m_*^2(x) = {g T^2(x) \over 24} \;.
\end{equation}
One sees that the condition $T \gg m_*$ is automatically satisfied
in the perturbative limit where $g^{-1} \gg 1.$

\subsection*{$\phi^3$ model}

\qquad Now we have the nonzero contributions to the self-energy 
not only from $\Pi_{\delta}$ (\ref{pi-delta-3-f}) but from $\Pi^{\gl}$  
(\ref{pi->-3-f}, \ref{pi-<-3-f}) as well. The real part of $\Pi^{+}$ 
which enters eq.~(\ref{mstar}) is given by the equation analogous 
to (\ref{Redelta}). Thus,
$$
{\rm Re} \Pi^{+}(X,m_* , {\bf 0})
= {1 \over 2 \pi i} {\rm P}
\int d\omega \; {\Pi^>(X,\omega , {\bf 0})
- \Pi^< (X,\omega , {\bf 0}) \over m_* - \omega} \;.
$$
Using eqs.~(\ref{pi->-3-eq}, \ref{pi-<-3-eq}) one finds
\begin{eqnarray}
{\rm Re} \Pi^{+}(X,m_* , {\bf 0}) &=& - {g^2 \over 16 \pi^2} \, m_*
\int_{2m_*}^{\infty} {d\omega \over \omega^2 -  m_*^2 } \,
\sqrt{1 - {4m_*^2 \over \omega^2}} \,
{1 \over \exp \big( {\beta \omega \over 2}\big) - 1 } \nonumber\\
&=& - {g^2 \over 32 \pi^2} \, I(\beta m_*) \;,
\end{eqnarray}
where
$$
I(a) \equiv \int_1^{\infty} {dx \over x^2 -  1/4 } \,
{\sqrt{1 - 1/x^2}  \over e^{ax} - 1 } \;.
$$
Thus, in the case of {\it global} equilibrium with $T \gg m_*$
the equation (\ref{mstar}) reads
\begin{equation}\label{eq-mass-3-eq}
m_*^2 + {g^2 \over 24} \, {T^2 \over m_*^2} 
- {g^2 \over 32 \pi^2} \, I(\beta m_*)  = 0 \;.
\end{equation}

One observes that
$$
I(a) < 
{1 \over a} \int_1^{\infty} {dx \over x^2 } = {1 \over a} \;.
$$
Thus, $I(\beta m_*) < T/m_*$ and consequently the absolute value of
the third term in l.h.s  of eq.~(\ref{eq-mass-3-eq}) is much smaller 
then that of the second one in the limit of $T \gg m_*$. Therefore, there 
is {\it no} real positive $m_*$ which satisfies eq. (\ref{eq-mass-3-eq}). 
It means that {\it the massive quasiparticles do not emerge in the massless 
$\phi^3$ model}. It is not surprising since the potential 
$V(\phi) = {g \over 3!}\phi^3$ from the lagrangian (\ref{lagran}) has
no minimum, even a local one. So, we do not consider the $\phi^3$ model 
any more and concentrate on the $\phi^4$ model.

\vspace{1cm}

Having the mass of quasiparticles we can determine their dispersion 
relation. Since all self-energies except $\Pi_{\delta}(X)$ vanish in the 
lowest order of the perturbative expansion, the mass-shell equation 
(\ref{mass1}) coincides with that one of the free quasiparticles 
(\ref{mass0}). Thus, we have explicitly shown that in the first order
of $g$, the $\phi^4$ model provides the system of free quasiparticles 
described by the transport equation of the Vlasov form i.e.
\begin{eqnarray}\label{trans-f0}
\Bigr[p^{\mu} \partial_{\mu} + 
{1 \over 2} \partial_{\mu} m_*^2(X) \partial^{\mu}_p \Bigl] 
f_0(X,{\bf p}) = 0 \;,
\end{eqnarray}
with the effective mass given by eq.~(\ref{eff-mass}).

\section{Higher order self-energy}

\qquad We discuss here the $g^2$ contributions to the self-energy in
the $\phi^4$ model, which are represented by graphs shown in Fig.~4. 
The bubble is again related to the effective mass. The contributions 
corresponding to the diagrams 4a and 4b can be easily computed. However, 
they are pure real and the only effect of these contributions is a higher 
order modification of the effective mass. Thus, we do not explicitly 
calculate these diagrams but instead we analyse the contribution 4c 
which provides a qualitatively new effect.

The graph from Fig.~4c gives the contour self-energy as
$$
\Pi_c(x,y) = {g^2 \over 6} \Delta_0(x,y) \Delta_0(y,x) \Delta_0(x,y) \;,
$$
and consequently
$$
\Pi^{\gl}(x,y) = {g^2 \over 6}
\Delta^{\gl}_0(x,y) \Delta^{\lg}_0(y,x) \Delta^{\gl}_0(x,y)\;. 
$$

Substituting $\Delta^{\gl}_0$ expressed through the distribution 
function $f_0$ as in eqs.~(\ref{sm-f0}) and (\ref{gr-f0}) one finds the 
self-energy $\Pi^{\gl}$ in the form 
\begin{eqnarray}\label{pi-<-4}
\Pi^<(X,p) &=& i{g^2\over 6} \int {d^3 k \over (2\pi )^3 2E_k} \,
{d^3 q \over (2\pi )^3 2E_q} \,
{d^3 r \over (2\pi )^3 2E_r} \, (2\pi)^4 
\delta^{(3)} ({\bf p} + {\bf q} - {\bf k} - {\bf r}) \nonumber\\
&\times&\Big[ \;\;\;\;\;
    \delta (p_0 - E_k - E_q - E_r) \, f_0^k \, f_0^{-q} \, f_0^r \nonumber\\
&&\;\;\;+\; \delta (p_0 - E_k - E_q + E_r) \, f_0^k \, f_0^{-q}\, (f_0^{-r} +1) \nonumber\\
&&\;\;\;+\; \delta (p_0 - E_k + E_q - E_r) \, f_0^k \, (f_0^q + 1) \, f_0^r  \nonumber\\
&&\;\;\;+\; \delta (p_0 - E_k + E_q + E_r) \, f_0^k \, (f_0^q +1) (f_0^{-r} +1) \nonumber\\
&&\;\;\;+\; \delta (p_0 + E_k - E_q - E_r) \, (f_0^{-k}+1) \, f_0^{-q} \, f_0^r \nonumber\\
&&\;\;\;+\; \delta (p_0 + E_k - E_q + E_r) \, (f_0^{-k}+1) \, f_0^{-q}\, (f_0^{-r} +1) \nonumber\\
&&\;\;\;+\; \delta (p_0 + E_k + E_q - E_r) \, (f_0^{-k}+1) \, (f_0^q + 1) \, f_0^r  \nonumber\\
&&\;\;\;+\; \delta (p_0 + E_k + E_q + E_r) \, (f_0^{-k}+1) \, (f_0^q +1) (f_0^{-r} +1)\Big]\;,
\end{eqnarray}
\begin{eqnarray}\label{pi->-4}
\Pi^>(X,p) &=& i{g^2\over 6} \int {d^3 k \over (2\pi )^3 2E_k} \,
{d^3 q \over (2\pi )^3 2E_q} \,
{d^3 r \over (2\pi )^3 2E_r} \, (2\pi)^4 
\delta^{(3)} ({\bf p} + {\bf q} - {\bf k} - {\bf r}) \nonumber\\
&\times&\Big[ \;\;\;\;\;
            \delta (p_0 + E_k + E_q + E_r) \, f_0^{-k} \, f_0^q \, f_0^{-r} \nonumber\\
&&\;\;\;+\; \delta (p_0 + E_k + E_q - E_r) \, f_0^{-k} \, f_0^q\, (f_0^r +1) \nonumber\\
&&\;\;\;+\; \delta (p_0 + E_k - E_q + E_r) \, f_0^{-k} \, (f_0^{-q} + 1) \, f_0^{-r}  \nonumber\\
&&\;\;\;+\; \delta (p_0 + E_k - E_q - E_r) \, f_0^{-k} \, (f_0^{-q} +1) (f_0^r +1) \nonumber\\
&&\;\;\;+\; \delta (p_0 - E_k + E_q + E_r) \, (f_0^k+1) \, f_0^q \, f_0^{-r} \nonumber\\
&&\;\;\;+\; \delta (p_0 - E_k + E_q - E_r) \, (f_0^k+1) \, f_0^q\, (f_0^r +1) \nonumber\\
&&\;\;\;+\; \delta (p_0 - E_k - E_q + E_r) \, (f_0^k+1) \, (f_0^{-q} + 1) \, f_0^{-r}  \nonumber\\
&&\;\;\;+\; \delta (p_0 - E_k - E_q - E_r) \, (f_0^k+1) \, (f_0^{-q} +1) (f_0^r +1)\Big]\;,
\end{eqnarray}
with
$$
f_0^k \equiv f_0(X,{\bf k}) \;,\;\;\;\;\;\;
f_0^{-k} \equiv f_0(X,- {\bf k}) \;.
$$
It is important to notice that in contrast to the similar expressions
of the $\phi^3$ model i.e. eqs.~(\ref{pi->-3-f}) and (\ref{pi-<-3-f}), 
the self-energies (\ref{pi->-4}) and (\ref{pi-<-4}) are nonzero not
only for the off-shell but for on-shell momentum $p$ as well. 
However, the number of terms from eq.~(\ref{pi->-4}) or (\ref{pi-<-4}) 
which contribute to $\Pi^{\gl}$ is reduced when $p^2 = m_*^2$. Indeed, 
eqs.~(\ref{pi-<-4}, \ref{pi->-4}) simplify in this case as
\begin{eqnarray}\label{pi-<-4-mshell}
\Theta(p_0)\,\Pi^<(X,p) &=& i{g^2\over 6} \int {d^3 k \over (2\pi )^3 2E_k} \,
{d^3 q \over (2\pi )^3 2E_q} \,
{d^3 r \over (2\pi )^3 2E_r} \, (2\pi)^4 
\delta^{(3)} ({\bf p} + {\bf q} - {\bf k} - {\bf r}) \nonumber\\
&\times&\Big[ \;\;\;\;\;
      \delta (p_0 - E_k - E_q + E_r) \, f_0^k \, f_0^{-q}\, (f_0^{-r} +1) \nonumber\\
&&\;\;\;+\; \delta (p_0 - E_k + E_q - E_r) \, f_0^k \, (f_0^q + 1) \, f_0^r  \nonumber\\
&&\;\;\;+\; \delta (p_0 + E_k - E_q - E_r) \, (f_0^{-k}+1) \, f_0^{-q} \, f_0^r \Big]\nonumber\\
&=& i{g^2\over 2} \int {d^3 k \over (2\pi )^3 2E_k} \,
{d^3 q \over (2\pi )^3 2E_q} \,
{d^3 r \over (2\pi )^3 2E_r} \, (2\pi)^4 
\delta^{(4)} (p + q - k - r) \nonumber\\
&\times& \, (f_0^q + 1) \, f_0^k  \, f_0^r  \;,
\end{eqnarray}
\begin{eqnarray}\label{pi->-4-mshell}
\Theta(p_0)\,\Pi^>(X,p) &=& i{g^2\over 6} \int {d^3 k \over (2\pi )^3 2E_k} \,
{d^3 q \over (2\pi )^3 2E_q} \,
{d^3 r \over (2\pi )^3 2E_r} \, (2\pi)^4 
\delta^{(3)} ({\bf p} + {\bf q} - {\bf k} - {\bf r}) \nonumber\\
&\times&\Big[ \;\;\;\;\;
      \delta (p_0 - E_k - E_q + E_r) \, (f_0^k+1) \, (f_0^{-q}+1)\, 
                                                   f_0^{-r} \nonumber\\
&&\;\;\;+\; \delta (p_0 - E_k + E_q - E_r) \,(f_0^k+1) \, f_0^q \, 
                                                   (f_0^r +1) \nonumber\\
&&\;\;\;+\; \delta (p_0 + E_k - E_q - E_r) \,f_0^{-k} \, (f_0^{-q}+1) \, 
                                                 (f_0^r+1) \Big]\nonumber\\
&=& i{g^2\over 2} \int {d^3 k \over (2\pi )^3 2E_k} \,
{d^3 q \over (2\pi )^3 2E_q} \,
{d^3 r \over (2\pi )^3 2E_r} \, (2\pi)^4 \delta^{(4)} (p + q - k - r) 
                                                                \nonumber\\
&\times& \, f_0^q \, (f_0^k+1)  \, (f_0^r +1) \;,
\end{eqnarray}
where $p_0 = \sqrt{m_*^2 + {\bf p}^2}$.

The self-energies $\Pi^{\gl}$ provide, through the equations analogous
to (\ref{Imdelta}) and (\ref{Redelta}), the ${\rm Re} \Pi^+$ and
${\rm Im} \Pi^+$ which enter the transport (\ref{trans1}) and mass-shell 
(\ref{mass1}) equations. The imaginary part of $\Pi^+$ is of particular 
interest. Due to the finite value of ${\rm Im} \Pi^+$, the spectral function 
(\ref{spec}) is no longer delta-like but it is of the Breit-Wigner shape. 
Thus, the quasiparticles are of finite life time. For the on-mass-shell 
momenta with $p_0 > 0$ the imaginary part of $\Pi^+$ equals
\begin{eqnarray}\label{Im-pi-mshell}
{\rm Im}\Pi^+(X,p)  &=&  {1 \over 2i} \, 
\Big(\Pi^>(X,p) - \Pi^<(X,p) \Big) \nonumber\\
&=& {g^2\over 4} \int {d^3 k \over (2\pi )^3 2E_k} \,
{d^3 q \over (2\pi )^3 2E_q} \,
{d^3 r \over (2\pi )^3 2E_r} \, (2\pi)^4 
\delta^{(4)} (p + q - k - r) \nonumber\\
&\times& \bigg( f_0^q \, (f_0^k+1)  \, (f_0^r +1) 
- (f_0^q + 1) \, f_0^k  \, f_0^r  \bigg) \;.
\end{eqnarray}
This function was computed for the equilibrium distribution in 
\cite{Par92,Wan96}. 

\section{Interacting quasiparticles}

\qquad In this chapter we discuss the dispersion relation of the 
interacting quasiparticles and then define the respective distribution 
function. 

Having the self-energies calculated in $g^2$ order we can determine
the quasiparticle dispersion relation in this order. As previously, the
quasiparticle mass is found as a solution of the equation (\ref{mstar}).
Therefore, the singular self-energy  $\Pi_{\delta}(X)$ as well as
${\rm Re} \Pi^{+}(X,m_* , {\bf 0})$ are included in $m_*^2(X)$. To avoid
the double counting, the expression 
$$
p^2 - m_*^2(X) + \Pi_{\delta}(X) + {\rm Re} \Pi^+(X,p)
$$
is everywhere replaced by
$$
p^2 - m_*^2(X) + {\rm Re} \widetilde\Pi^+(X,p) \;,
$$
with
$$
{\rm Re}\widetilde\Pi^+(X,p) \equiv + {\rm Re}\Pi^+(X,p) 
- \Theta(p_0) \, {\rm Re} \Pi^+(X,m_* , {\bf 0})
- \Theta(-p_0) \,{\rm Re} \Pi^+(X,-m_* , {\bf 0}) \;.
$$

The dispersion relation is given by the equation
\begin{equation}\label{dis-re}
p^2 - m_*^2(X) + {\rm Re} \widetilde\Pi^+(X,p) = 0. \;,
\end{equation}
but according to eq.~(\ref{spec}), which determines the spectral
function, the quasiparticles are of the finite width and the relation
(\ref{dis-re}) gives only the most probable quasiparticle four-momentum.
We call the four-momenta, which satisfy the relation (\ref{dis-re}), as
`on-mass-shell', however one should keep in mind that the meaning of this 
term differs for the finite and zero width quasiparticles.

Eq.~(\ref{dis-re}) can be easily solved if ${\rm Re}\widetilde\Pi^+$ 
provides only a small correction to the free quasiparticle dispersion 
relation. Then, one finds the on-mass-shell momentum as 
$p^{\pm} = (\pm E^{\pm}_p , {\bf p})$ with
\begin{equation}\label{energy}
E^{\pm}_p = \sqrt{m_*^2(X) + {\bf p}^2 +
{\rm Re} \widetilde\Pi^+(X,\pm\sqrt{m_*^2(X) + {\bf p}^2}, {\bf p})} \;.
\end{equation}

The distribution function $f(X,p)$ of the interacting quasiparticles is
defined in a way analogous to eq.~(\ref{def1-f0}) i.e.
\begin{eqnarray}\label{def-f}
\Theta (p_0) i \Delta^<(X,p) = \Theta (p_0)\; A(X,p) \; f(X,p) \;,
\end{eqnarray}
where $A(X,p)$ is the spectral function (\ref{spec}). In contrast to 
the case of free quasiparticles, $f(X,p)$ depends not on the three-vector 
${\bf p}$ but on the four-vector $p$. Due to the identities 
$$
\Delta^<(X,p) = \Delta^>(X,-p) \;,\;\;\;\;\;
A(X,p) = i\Delta^>(X,p) - i\Delta^<(X,p) \;,
$$
we have
\begin{eqnarray}\label{gr-f}
i \Delta^>(X,p)  =  \Theta (p_0) \; A(X,p) \; \big( f(X,p) + 1 \big)
- \Theta (-p_0) \; A(X,p) \; f(X,-p) \;,
\end{eqnarray}
\begin{eqnarray}\label{sm-f}
i \Delta^<(X,p)  = \Theta (p_0) \; A(X,p) \; f(X,p) 
- \Theta (-p_0) \; A(X,p) \; \big( f(X,-p) + 1 \big) \;.
\end{eqnarray}

There is a very important property of $\Delta^{\gl}$ expressed in the
from (\ref{gr-f}, \ref{sm-f}). Namely, if the Green functions 
$\Delta^{\gl}$ satisfy the transport equation (\ref{trans1}) and 
the spectral function solves the equation (\ref{mass-spec}), the 
mass-shell equation of $\Delta^{\gl}$, i.e. eq.~(\ref{mass1}), is satisfied 
{\it automatically} in the 0-th order of the gradient expansion. 
Let us derive this result.  

The transport and mass-shell equations (\ref{trans1}, \ref{mass1}) with 
the gradient terms neglected read
$$ 
0 =  \Pi^>(X,p) \Delta^<(X,p) - \Pi^< (X,p) \Delta^> (X,p) \;,
$$
$$
\Bigr[ p^2 - m_*^2(X) + {\rm Re}\widetilde\Pi^+(X,p) 
\Bigl] \Delta^{\gl }(X,p)  
= - \Pi^{\gl}(X,p) {\rm Re} \Delta^+(X,p) \;.
$$
Substituting $\Delta^{\gl}$ in the form (\ref{gr-f}, \ref{sm-f})
into the first equation and taking only the terms corresponding to
$p_0 > 0$, the equation is
\begin{equation}\label{tran-0-f} 
0 = A(x,p)
\bigg[ \Pi^>(X,p) \, f(X,p) - \Pi^< (X,p) \, \Big( f(X,p)+1\Big) \bigg] \;.
\end{equation}
Now we substitute $\Delta^<$ given by (\ref{sm-f}) into the mass-shell
equation and get
\begin{equation}\label{mass-0-f-1}
\Bigr[ p^2 - m_*^2(X) + {\rm Re} \widetilde\Pi^+(X,p) 
\Bigl] \, A(X,p) \, f(X,p)  
= - \Pi^<(X,p) {\rm Re} \Delta^+(X,p) \;,
\end{equation}
where $p_0$ is assumed to be positive. Using the spectral function equation 
(\ref{mass-spec}), the equation (\ref{mass-0-f-1}) is manipulated to the form
\begin{equation}\label{mass-0-f-2} 
{\rm Re}\Delta^+(x,p) \;
\bigg[ \Pi^>(X,p) \, f(X,p) - \Pi^< (X,p) \, \Big( f(X,p)+1\Big) \bigg] = 0 \;.
\end{equation}
One sees that if $f$ solves eq.~(\ref{tran-0-f}), it automatically
satisfies eq.~(\ref{mass-0-f-2}). Similar considerations can be easily 
repeated for $p_0 < 0$ and then for $\Delta^<$ with $p_0 > 0$ and $p_0 < 0$. 

Let us observe that the quasiparticles studied in this paper are
{\it narrow}. Indeed, the effective mass (\ref{eff-mass}) is of order 
$g^{1/2}$ while the width of the Breit-Wigner distribution given by 
eq.~(\ref{width}) is proportional to $g$ in power at least $3/2$. Thus, 
the width of quasiparticles is much smaller than their mass ($g$ is obviously 
assumed to be small). Due to this property we will often refer to the case
of zero-width quasiparticles or on-mass-shell momenta.

\section{Transport equation}

\qquad The distribution function $f$ satisfies the transport equation
which can be obtained from eq.~(\ref{trans1}) for $\Delta^>$ or $\Delta^<$. 
After using eq.~(\ref{trans-spec}) one finds
\begin{eqnarray}\label{trans-f}
A(X,p)\,\Big\{ p^2 - m_*^2(X) &+& {\rm Re}\widetilde\Pi^+(X,p), 
\, f(X,p) \Big\} \nonumber \\
&=& i A(X,p) \,\Big( \Pi^>(X,p) \, f(X,p) -
      \Pi^< (X,p) \, \big(f(X,p)+1) \Big) \nonumber \\
&+& if(X,p) \, \Big\{ \Pi^>(X,p),\, {\rm Re} \Delta^+(X,p) \Big\} \nonumber \\
&-& i\big(f(X,p)+1\big)\,\Big\{ \Pi^<(X,p),\,{\rm Re} \Delta^+(X,p) \Big\}\;,
\end{eqnarray}
where $p_0 > 0$. We have also used here the following property of the 
Poisson-like brackets:
$$
\big\{A ,\, B\,C \big\} = \big\{A ,\, B \big\}\,C 
+ \big\{A ,\, C \big\}\,B \;.
$$
Since eq.~(\ref{trans-f}) is one of the main results of this paper we discuss 
it in detail.

The left-hand-side of eq.~(\ref{trans-f}) is a straightforward generalization
of the drift term of the standard relativistic transport equation. Computing
the Poisson-like bracket and imposing the mass-shell constraint (\ref{dis-re})
one finds the familiar structure 
\begin{eqnarray*}
{1 \over 2} \Theta(p_0) \,
\Big\{ p^2 - m_*^2(X) &+& {\rm Re}\widetilde\Pi^+(X,p), \, f(X,p) \Big\} \\
&=& E^+_p \Big({\partial \over \partial t} + {\bf v} \nabla \Big) f(X,p)
+ \nabla V(X) \, \nabla_p f(X,p)
\end{eqnarray*}
where
$$
V(X) \equiv m_*^2(X) - {\rm Re}\widetilde\Pi^+(X,p) \;,
$$
and the velocity ${\bf v}$ equals $\partial E^+_p /\partial {\bf p}$
with the energy $E^+_p$ given by eq.~(\ref{energy}).

Let us now analyse the right-hand-side of eq.~(\ref{trans-f}). Since the 
quasiparticles of interest are narrow, we take into account only 
those terms contributing to the self-energies $\Pi^{\gl}$ which are nonzero 
for the on-mass-shell momenta. The other terms are negligibly small. Then, 
$\Pi^{\gl}$ from the transport equation (\ref{trans-f}) are given by 
the formulas analogous to eqs.~(\ref{pi-<-4-mshell},\ref{pi->-4-mshell}) 
with $f$ instead of $f_0$. Consequently,
\begin{eqnarray}\label{pi-<-4-mshell-f}
\Theta(p_0)\,\Pi^<(X,p) &=& 
i{g^2\over 2} \int {d^4 k A^+_k\over (2\pi )^4 } \,
{d^4 q A^+_q \over (2\pi )^4} \,
{d^4 r A^+_r \over (2\pi )^4} \, (2\pi)^4 
\delta^{(4)} (p + q - k - r) \nonumber\\
&\times& \, (f^q + 1) \, f^k  \, f^r  \;,
\end{eqnarray}
\begin{eqnarray}\label{pi->-4-mshell-f}
\Theta(p_0)\,\Pi^>(X,p) 
&=& i{g^2\over 2} \int {d^4 k A^+_k \over (2\pi )^4} \,
{d^4 q A^+_q \over (2\pi )^4} \,
{d^4 r A^+_r \over (2\pi )^4} \, 
(2\pi)^4 \delta^{(4)} (p + q - k - r) \nonumber\\
&\times& \, f^q \, (f^k+1)  \, (f^r +1) \;,
\end{eqnarray}
where
$$
A^+_k \equiv \Theta(k_0) \; A(X,k) \;.
$$
One sees that in the limit of zero-width quasiparticles
$$
{d^4 k A^+_k\over (2\pi )^4 } \rightarrow {d^3 k \over (2\pi )^3 2E_k} \;.
$$

The first term in r.h.s of the transport equation (\ref{trans-f}) is very
similar to the standard collision term of the relativistic transport
equation \cite{Gro80}. Indeed,
\begin{eqnarray}\label{coll}
i\Big( \Pi^>(X,p) \, f(X,p) &-& \Pi^< (X,p) \, \big(f(X,p)+1) \Big) \\ 
&=& {g^2\over 2} \int {d^4 k A^+_k \over (2\pi )^4} \,
{d^4 q A^+_q \over (2\pi )^4} \,
{d^4 r A^+_r \over (2\pi )^4} \, 
(2\pi)^4 \delta^{(4)} (p + q - k - r) \nonumber\\
&\times& \bigg( (f^p + 1) \;(f^q + 1) \, f^k  \, f^r - 
f^p\, f^q \, (f^k+1)  \, (f^r +1) \bigg)\,. \nonumber
\end{eqnarray}

The last two terms from r.h.s of eq.~(\ref{trans-f}), analogous to those 
found long time ago in \cite{Kad62,Bez72}, are absent in the usual transport 
equation. We are going to show that in the local equilibrium, when the 
collision term (\ref{coll}) vanish, we reproduce the standard collisionless 
equation if the four-momentum is on mass-shell.

As well known \cite{Gro80}, the standard collision term, which emerges 
from (\ref{coll}) when the quasiparticle width tends to zero 
(cf. eq.~(\ref{spec-free})), vanishes for the local equilibrium distribution 
function of the form (\ref{equi}). Following \cite{Gro80}, one easily
shows that the collision term (\ref{coll}) also vanishes for the 
distribution function (\ref{equi}) with the particle momentum $p$ no
longer constrained by the mass-shell condition.

One observes that in the local equilibrium the collisional self-energies
can be written as
$$
\Pi^>(X,p) = 2i {\rm Im}\Pi^+(X,p) \, 
\big(f^{\rm eq}(X,p) + 1\big) \;,
$$
and 
$$
\Pi^<(X,p) = 2i {\rm Im}\Pi^+(X,p) \, f^{\rm eq}(X,p) \,.
$$
The transport equation (\ref{trans-f}) then simplifies to
\begin{eqnarray*}
A(X,p)\,\Big\{ p^2 - m_*^2(X) &+& {\rm Re}\widetilde\Pi^+(X,p), 
\, f^{\rm eq}(X,p) \Big\} \\
&=&  2 {\rm Im}\Pi^+(X,p) \,
\Big\{f^{\rm eq}(X,p),\,{\rm Re} \Delta^+(X,p) \Big\} \,.
\end{eqnarray*}
Using eqs.~(\ref{Re-pm}) and (\ref{spec}) one manipulates this equation
to the form
\begin{eqnarray}\label{trans-vlas}
{\rm Im}\Pi^+(X,p)\,\Big\{ p^2 &-& m_*^2(X) 
+ {\rm Re} \widetilde\Pi^+(X,p), \, f^{\rm eq}(X,p) \Big\} \\ 
&=& \Big( p^2 - m_*^2(X) + {\rm Re}\widetilde\Pi^+(X,p) \Big) 
\Big\{ {\rm Im}\Pi^+(X,p) ,\, f^{\rm eq}(X,p) \Big\} \,.\nonumber
\end{eqnarray}
As seen, the term on the r.h.s drops down for the on-mass-shell momenta 
and then we reproduce the usual Vlasov equation i.e.
\begin{eqnarray}\label{trans-vlas-stand}
\Big\{ p^2 - m_*^2(X) + {\rm Re} \widetilde\Pi^+(X,p), 
\, f^{\rm eq}(X,p) \Big\} = 0 \,.
\end{eqnarray}

The role of the two unusual terms from r.h.s of the transport equation 
(\ref{trans-f}) beyond the local equilibrium is rather unclear and needs 
further studies.

\section{Summary and concluding remarks}

\qquad We have discussed in this paper the nonequilibrium features 
of the massless fields. The derivation of the kinetic equation in such 
a case faces serious difficulties because there is no natural length scale 
over which the system inhomogeneities can be integrated over. As known 
the transport theory deals with the quantities averaged over an elementary 
phase-space cell of the minimal size given by the particle Compton 
wavelength.

The fields with the zero bare mass usually gain an effective mass due 
to the self-interaction. Therefore, we have introduced the auxiliary
mass term in the lagrangian and then, the transport theory has been 
derived in a way very similar to the earlier studied \cite{Mro90,Mro94}
case of massive fields. However, due to the position dependence of the 
effective mass, the limit of the noninteracting quasiparticles corresponds 
to the Vlasov rather than the free particle case. The smallness of the 
effective mass has also forced us to take into  account some extra gradient
terms which are usually neglected in the transport equation.

We have considered in detail the $\phi^3$ and $\phi^4$ models which
appear to be very different from each other. In the $\phi^4$ model 
the effective mass is generated in the lowest nontrivial order of
the perturbative expansion. In contrast, the massive quasiparticles 
do not emerge in the $\phi^3$ model and most probably there is no transport 
limit of this model which, as well known, is anyhow ill defined.

Within the $\phi^4$ model we have derived the transport equation for
the finite width quasiparticles. The distribution function has been defined
in such a way that the mass-shell constraint is automatically satisfied 
(in the gradient zeroth order). We have found except the mean-field and 
collision terms the specific ones which are absent in the standard transport
equation. However, in the case of local equilibrium we have been able to 
reproduce the usual collisionless equation if the four-mementum is on the 
mass shell.

The massless fields play a crucial role in the gauge theories such as Quantum
Chromodynamics. We believe that the methods developed in this study will be 
useful in the discussion of the transport theory of quarks and gluons. 
Till know only the mean-field limit of such a theory is well understood
\cite{Elz89,Bla93}. In spite of some efforts \cite{Sel93,Gei96}, the 
systematic derivation of the collision terms is still missing. The problem 
of the off-mass-shell propagation, which on phenomenological level has been
proven to play a very important role \cite{Gei95}, is also unsettled.

\section*{Acknowledgements}

\qquad I am very grateful to Heribert Weigert for suggesting me to introduce
the auxiliary mass term in the lagrangian to study the fields with the
zero bare mass. This work was partially supported by the Polish Committee 
of Scientific Research under Grant No. 2 P03B 195 09.

\section*{Appendix}

\qquad We discuss here the quasiparticle approximation for the system of
noninteracting fields. To simplify the discussion the bare mass $m$
is assumed to be nonzero or equivalently $m_*$ is treated as a constant. 
The transport equation and the mass-shell constraint read  
\begin{equation}\label{eq-trans}
p \cdot \partial \, \Delta^{\gl}_0(X,p) = 0 \;,   
\end{equation}
\begin{equation}\label{eq-mass}
[{1 \over 4} \partial ^2 
- p^2 + m^2] \: \Delta^{\gl}_0(X,p) = 0 \;.
\end{equation}
These equations, which directly follow from the field equation of motion 
(\ref{motion}) with $m = m_*$, are exact in the case of the massive free 
fields - the gradient expansion is not needed to derive them.

The mass-shell constraint (\ref{eq-mass}) shows that the function 
$\Delta^{\gl}_0(X,p)$ is indeed nonzero for the off-shell momenta i.e.
$\Delta^{\gl}_0(X,p)\not= 0$ for $p^2 \not= m^2$. This result looks 
surprising if one keeps in mind that the field, which solves the equation 
of motion (\ref{motion}), is in a sense on mass-shell. The field is the sum 
of the plane waves 
\begin{equation}\label{field-sol}
\phi(x) =   \int {d^3k \over \sqrt{(2\pi )^3 2 \omega }} 
\Big( e^{-ikx} \; a({\bf k}) + e^{ikx} \; a^*({\bf k}) \Big) \;, 
\end{equation}
where $k \equiv (\omega, {\bf k})$ with 
$\omega \equiv \sqrt{{\bf k}^2 + m^2}$. Thus, $k^2 = m^2$. Substituting 
the field (\ref{field-sol}) into the $\Delta^{\gl}(X,p)$ definition,
one finds that the off-shell contribution to $\Delta^{\gl}_0(X,p)$ comes 
from the interference of the positive and negative energy parts present in 
eq.~(\ref{field-sol}). Let us consider when such a contribution can 
be neglected.

One easily shows that the transport equation (\ref{eq-trans}) is solved by 
the function which depends on the four-position $X=(t,{\bf x})$ only through 
${\bf x} - {\bf v}t$ i.e.
$$
\Delta^{\gl}_0(X,p) = F({\bf x} - {\bf v}t,p) \;.
$$
The quasiparticle condition (\ref{quasipar}) applied to the function $F$ reads
$$
\Big\vert F({\bf x} - {\bf v}t,p) \Big\vert  \gg { 1\over m^2} 
\Big\vert  (v_i v_j - \delta_{ij}) \;
{\partial^2 F({\bf x} - {\bf v}t,p) \over 
\partial ({\bf x} - {\bf v}t)_i \partial ({\bf x} - {\bf v}t)_j} 
\Big\vert \;.
$$
If this condition is satisfied for every ${\bf x}$ at a given moment of time, 
say $t_0$, it is satisfied at {\it any} time. In other words, if the initial 
condition at $t_0$ is sufficiently homogeneous that the quasiparticle 
approximation can be applied, then this approximation is applicable at any 
time - the system remains homogeneous.

The question arises whether $\Delta^{\gl}_0$, which simultaneously solves 
the transport (\ref{eq-trans}) and mass-shell (\ref{eq-mass}) equations, 
{\it can} satisfy the quasiparticle condition. We introduce the Fourier 
transformed function $\widetilde\Delta^{\gl}_0(Q,p)$ defined as 
\begin{equation}\label{Wigner-def-q}
\widetilde\Delta^{\gl}_0(Q,p) 
\buildrel \rm def \over = \int d^4X \; 
e^{iQ \cdot X} \;  \Delta^{\gl}_0(X,p) \;. 
\end{equation}
The equations corresponding to (\ref{eq-trans}) and (\ref{eq-mass}), 
respectively, read
$$
p \cdot Q \; \widetilde\Delta^{\gl}_0(Q,p) = 0 \;, 
$$
$$
[-{1 \over 4} Q^2 - p^2 + m^2] \: \widetilde\Delta^{\gl}_0(Q,p) = 0 \;.
$$
They are both solved by
\begin{equation}\label{solution}
i\widetilde\Delta^{\gl}_0(Q,p)  = \delta ( p \cdot Q) \;
\delta (-{1 \over 4} Q^2 - p^2 + m^2) \; A(Q,p) \;, 
\end{equation}
with $A(Q,p)$ controlled by the initial condition. Since 
$i\Delta^{\gl}_0(X,p)$ is real, $A(Q,p)$ has the property 
\begin{equation}\label{prop-A}
A(Q,p) = A^*(-Q,p) \;.
\end{equation}

The solution of the equations (\ref{eq-trans}) and (\ref{eq-mass}) 
satisfies the quasiparticle condition (\ref{quasipar}) when
\begin{equation}\label{quasipar-A}
\Big\vert A(Q,p) \Big\vert \gg 
\Big\vert {Q^2 \over m^2} \, A(Q,p) \Big\vert \;,
\end{equation}
or equivalently $A(Q,p) \not= 0$ only for $Q^2 \ll m^2$.

It is instructive to consider the explicit solution of eqs.~(\ref{eq-trans}) 
and (\ref{eq-mass}) in $1 + 1$ dimensions. Using (\ref{solution}) we get
\begin{eqnarray}\label{sol-1+1-1}
i\Delta^{\gl}_0(X,p) &=& \int {d^2Q \over (2\pi )^2}\;
e^{ -iQ \cdot X} \; \delta ( p \cdot Q ) \;
\delta (-{1 \over 4} q^2 - p^2 + m^2) \; A(Q,p) \;, 
\nonumber\\
&=& \Big[ \Theta(-p^2 ) + \Theta( p^2 - m^2) \Big] \; 
{1 \over (2\pi )^2 \vert p^2 \vert} \; \sqrt{p^2 \over p^2 - m^2}
\nonumber\\
&\times&
\Big( e^{- i\tilde Q X} \; A(\tilde Q,p) 
+ e^{ i\tilde Q X} \; A(-\tilde Q,p) \Big) \;,
\end{eqnarray}
where $\tilde Q$ denotes the two-vector 
$$
\tilde Q \equiv 2 \vert p_0 \vert \sqrt{p^2 - m^2 \over p^2 } \;
\Big( {p_1 \over p_0} \, ,  1 \Big) \;.
$$
Keeping in mind the property (\ref{prop-A}) the solution (\ref{sol-1+1-1})
can be rewritten as
\begin{eqnarray}\label{sol-1+1-2}
i\Delta^{\gl}_0(X,p) = \Big[ \Theta(-p^2 ) + \Theta( p^2 - m^2) \Big] \; 
\Big(f(p) \, \sin(\tilde Q X) + g(p) \, \cos (\tilde Q X) \Big) \;,
\end{eqnarray}
where $f(p)$ and $g(p)$ are the real functions of $p$ determined by 
the initial condition.

The quasiparticle condition (\ref{quasipar}) is satisfied by 
(\ref{sol-1+1-2}) if
$$
\vert \tilde Q^2 \vert = 4 \vert  p^2 - m^2 \vert  \ll m^2 \;.
$$
One also sees that $\Delta^{\gl}_0(X,p) \sim \delta (p^2 -m^2)$ only for 
$\tilde Q = 0$. In other words, the function $\Delta^{\gl}_0(X,p)$ is strictly 
zero for the off-mass-shell momenta when the system is exactly homogeneous.
If we are interested in the weakly nonhomogeneous systems, the functions are 
nonzero for $p^2 > m^2$. Equivalently, if $p^2 \cong m^2$ then 
$p^2 > m^2$ but not $p^2 < m^2$. 

The properties of the function $\Delta^{\gl}_0$ in $1+1$ dimension 
can be trivially generalized to $3+1$ case showing the limitations
of the quasiparticle approximation.

\newpage
\vspace{1cm}
\begin{center}
{\bf Figure Captions}
\end{center}
\vspace{0.3cm}

\noindent
{\bf Fig. 1.} 
The contour along the time axis for an evaluation of the
operator expectation values.

\vspace{0.5cm}

\noindent
{\bf Fig. 2.} 
The lowest-order diagrams of the self-energy in the $\phi^4$ model.
The bubble in a) denotes the additional interaction due to the effective mass.

\vspace{0.5cm}

\noindent
{\bf Fig. 3.} 
The lowest-order diagrams of the self-energy in the $\phi^3$ model.
The bubble in a) denotes the additional interaction due to the effective mass.

\vspace{0.5cm}
 
\noindent
{\bf Fig. 4.} 
The second-order diagrams of the self-energy in the $\phi^4$ model.

\end{document}